\documentclass{amsart}
\usepackage{amssymb}
\usepackage{latexsym}

\date{\today}

\newcommand{\Z}{{\mathbb Z}}
\newcommand{\R}{{\mathbb R}}

\newcommand{\T}{{\mathbb T}}

\newtheorem{theorem}{Theorem}

\newtheorem{coro}{Corollary}

\newcounter{smalllist}

\begin{document}
\title[Positive Lyapunov Exponents for the Doubling Map]{Almost Everywhere Positivity of the Lyapunov
Exponent for the Doubling Map}
\author[D.\ Damanik, R.\ Killip]{David Damanik and Rowan Killip}
\address{Mathematics 253--37, California Institute of
Technology, Pasadena, CA 91125, USA}
\email{damanik@its.caltech.edu}
\address{Department of Mathematics, University of California, Los Angeles, CA 90055, USA}
\email{killip@math.ucla.edu}

\thanks{D.\ D.\ was supported in part by NSF grant DMS--0227289}

\begin{abstract}
We show that discrete one-dimensional Schr\"odinger operators on
the half-line with ergodic potentials generated by the doubling
map on the circle, $V_\theta(n) = f(2^n \theta)$, may be realized
as the half-line restrictions of a non-deterministic family of
whole-line operators. As a consequence, the Lyapunov exponent is
almost everywhere positive and the absolutely continuous spectrum
is almost surely empty.
\end{abstract}
\maketitle

Consider discrete Schr\"odinger operators
\begin{equation}\label{oper}
[H_\theta \phi](n) = \phi(n+1) + \phi(n-1) + \lambda f(2^n \theta)
\phi(n)
\end{equation}
on $\ell^2(\Z_+)$, $\Z_+ = \{1,2,\ldots\}$, with a Dirichlet
boundary condition, $\phi(0) = 0$. Here, $\lambda>0$ denotes the
coupling constant, $\theta \in \T = \R/\Z$, and $f: \T \to \R$ is
measurable, bounded, and non-constant.

Since the doubling map, $T : \T \to \T$, $\theta \mapsto 2\theta
\mod 1$, is ergodic with respect to Lebesgue measure on $\T$, the
Lyapunov exponent exists for every energy $E$, that is, there is a
function $\gamma : \R \to [0,\infty)$ such that
$$
\gamma(E) = \lim_{n \to \infty} \frac1n \log \| M(n,E,\theta) \|
\text{ for almost every } \theta,
$$
where $M(n,E,\theta)$ is the transfer matrix from $1$ to $n$ for
the operator $H_\theta$ at energy $E \in \R$, that is,
$$
M(n,E,\theta) = \left( \begin{array}{cc} E - \lambda f(2^n \theta)
& -1 \\ 1 & 0
\end{array} \right) \times \cdots \times \left( \begin{array}{cc} E - \lambda f(2 \theta) & -1 \\ 1 & 0
\end{array} \right).
$$

It is expected that $\gamma (E) > 0$ for every $E$ and that
$H_\theta$ has pure point spectrum with exponentially decaying
eigenfunctions for almost every $\theta$. The reason for this is
that the potentials are strongly mixing and similar to random
potentials (generated by i.i.d.\ random variables) for which these
results were established in the 1980's. There is a huge literature
on this subject; we refer the reader to \cite{cl,pf} for
detailed references.

However, very few results are known for the operators $H_\theta$ and these
are limited to the case of small coupling: Chulaevsky
and Spencer, \cite{cs}, proved an asymptotic formula for
$\gamma(E)$, in the limit $\lambda \to 0$. This implied positivity
of $\gamma(E)$ for all $0 < |E| < 2$ and $\lambda$ small,
which in turn was used by Bourgain and Schlag, \cite{bs}, to prove
that for $\lambda<\lambda_0$ and almost every $\theta$,
$H_\theta$ has pure point spectrum in $0 < \delta \le |E| \le 2 - \delta$ with
exponentially decaying eigenfunctions.  This is usually referred to as Anderson localization.

The results just described require a regularity assumption on $f$ (e.g., H\"older continuity).
Beyond small coupling, no spectral results have been established. In particular, it was not
known, for any $\lambda > 0$, whether there can be any
absolutely continuous spectrum (for a.e.\ $\theta$).\footnote{It
should be possible to exclude absolutely continuous spectrum using
Kotani's support theorem \cite{k2}; see \cite{cs,s2} for remarks
indicating this possibility and \cite{kks} for related
applications of the support theorem. However, this would not give
absence of absolutely continuous spectrum in full generality, that
is, this approach would only work for some $f$'s and some values
of $\lambda$.} In this note, we show that there is none.

\begin{theorem}\label{main}
Suppose that $\lambda > 0$ and $f$ is measurable, bounded, and
non-constant. Then, the Lyapunov exponent $\gamma(E)$ is positive
for almost every $E \in \R$ and the absolutely continuous spectrum
of $H_\theta$ is empty for almost every $\theta \in \T$.
\end{theorem}

\noindent\textit{Remark.} Note that if the binary expansion of
$\theta$ is periodic, the absolutely continuous spectrum of
$H_\theta$ is not empty so that the second statement cannot be
improved.

\medskip

As we mentioned above, it is expected that $\gamma(E)$ is strictly
positive for \textit{every} energy $E$. Once this can be shown,
the next natural step will then be to prove Anderson localization,
using, for example, methods from \cite{bs}. Since such a result is
currently out of reach, we note that, by general principles, the
almost everywhere positivity of $\gamma(E)$ yields the following
consequence in terms of a localization result.

Given $\alpha \in [0,\pi)$, let $H_{\theta,\alpha}$ denote the
operator which acts on $\ell^2(\Z_+)$ as in \eqref{oper}, but with
$\phi(0)$ given by $\cos (\alpha) \phi(0) + \sin (\alpha) \phi(1)
= 0$. Thus, the original operator (with a Dirichlet boundary
condition) corresponds to $\alpha = 0$. By
\cite[Theorem~13.4]{pf}, Theorem~\ref{main} implies the following:

\begin{coro}
Suppose that $\lambda > 0$ and $f$ is measurable, bounded, and
non-constant. Then, for almost every $\alpha \in [0,\pi)$ and
almost every $\theta \in \T$, the operator $H_{\theta,\alpha}$
has pure point spectrum, and all eigenfunctions decay
exponentially at infinity.
\end{coro}

It is interesting to note that there are two possible viewpoints
and both have been used successfully. The work of Chulaevsky and
Spencer showed that methods from the random case (specifically an
approach developed by Figotin and Pastur \cite{pf}) could be extended
to establish the asymptotic formula and positivity for the
Lyapunov exponent. On
the other hand, Anderson localization was then proven by Bourgain
and Schlag by adapting methods originally developed for models
with very little randomness, namely, with underlying dynamics
given by the shift \cite{bg,gs,j} and the skew-shift on the torus
\cite{bgs}.

The main ingredient in the proof of Theorem~\ref{main} will be a
result of Kotani, \cite{k} (see \cite{s} for an adaptation to the
discrete case), which shows that for whole-line operators with
non-deterministic potentials, the Lyapunov exponent is almost
everywhere positive and, consequently, the absolutely continuous spectrum is
almost surely empty. A potential is non-deterministic if it is not
determined uniquely by its restriction to a half-line. Thus, the
natural strategy will be to define a non-deterministic family of
whole-line operators whose restrictions to the right half-line
yield the family $\{H_\theta\}$. Note that the Lyapunov exponents
for the two families are the same and that the half-line operators
cannot have absolutely continuous spectrum if the whole-line
operators do not have any.

\begin{proof}[Proof of Theorem~\ref{main}.]
The first step is to conjugate the doubling map $T$ to a symbolic
shift via the binary expansion. Let $\Omega_+ = \{0,1\}^{\Z_+}$
and define $D : \Omega_+ \to \T$ by $D(\omega) = \sum_{n =
1}^\infty \omega_n 2^{-n}$. The shift transformation, $S :
\Omega_+ \to \Omega_+$, is given by $(S\omega)_n = \omega_{n+1}$.
Clearly, $D \circ S = T \circ D$.

Next we introduce a family of whole-line operators as follows. Let
$\Omega = \{0,1\}^\Z$ and define, for $\omega \in \Omega$, the
operator
$$
[H_\omega \phi](n) = \phi(n+1) + \phi(n-1) + V_\omega(n) \phi(n)
$$
in $\ell^2(\Z)$, where
$$
V_\omega(n) = f[D( \{ \omega_n, \omega_{n+1}, \omega_{n+2}, \ldots
\} )].
$$
The family $\{H_\omega\}_{\omega \in \Omega}$ is non-deterministic
since $V_\omega$ restricted to $\Z_+$ only depends on
$\{\omega_n\}_{n \ge 1}$ and hence, by non-constancy of $f$, we
cannot determine the values of $V_\omega(n)$ for $n \le 0$
uniquely from the knowledge of $V_\omega(n)$ for $n \ge 1$. It
follows from \cite{k,s} that the Lyapunov exponent for
$\{H_\omega\}_{\omega \in \Omega}$ is almost everywhere positive
and $\sigma_{{\rm ac}}(H_\omega)$ is empty for almost every
$\omega \in \Omega$ with respect to the
$(\frac{1}{2},\frac{1}{2})$-Bernoulli measure on $\Omega$.

Finally, let us consider the restrictions of $H_\omega$ to
$\ell^2(\Z_+)$, that is, let $H_\omega^+ = E^* H_\omega E$, where
$E : \ell^2(\Z_+) \to \ell^2(\Z)$ is the natural embedding.
Observe that $H_\omega^+ = H_\theta$, where $\theta =
D(\{\omega_1, \omega_2, \omega_2, \ldots\})$. This immediately
implies the statement on the positivity of the Lyapunov exponent
for the family $\{H_\theta\}_{\theta \in \T}$. As finite-rank
perturbations preserve absolutely continuous spectrum,
$\sigma_{{\rm ac}}(H_\omega^+) \subseteq \sigma_{{\rm
ac}}(H_\omega)$ for every $\omega \in \Omega$. This proves that
$\sigma_{{\rm ac}}(H_\omega^+) = \emptyset$ for almost every
$\omega \in \Omega$.
\end{proof}

\noindent\textit{Remark.}  The proof relies on only two properties of the doubling map:
it is not one-to-one and it can be extended to a dynamical system over $\Z$.  Consequently,
Theorem~\ref{main} extends to other models with these properties. For example,
$\theta \mapsto m\theta \mod 1$ for any integer $m \ge 2$.

\medskip

\noindent\textit{Acknowledgments.} We thank Svetlana Jitomirskaya
and Barry Simon for useful conversations.

\end{document}